\documentclass[aps,floatfix,superscriptaddress,showpacs,twocolumn]{revtex4}

\usepackage{epsfig}
\usepackage{amsmath,amssymb}
\usepackage{graphicx,psfrag}
\usepackage{dcolumn}
\usepackage{bm}
\usepackage{slashed}
\usepackage{color}
\usepackage{placeins}

\usepackage{comment}

\newcommand{\beq}{\begin{equation}}
\newcommand{\eeq}{\end{equation}}
\newcommand{\bea}{\begin{eqnarray}}
\newcommand{\eea}{\end{eqnarray}}

\newcommand{\nn}{\nonumber\\}

\newcommand\fig[1]     {Fig.\,{\ref{#1}}}

\def\eq#1{(\ref{#1})}
\def\s0#1#2{\mbox{\small{$ \frac{#1}{#2} $}}}
\def\0#1#2{\frac{#1}{#2}}

\def\mr#1{{\mathrm{#1}}}

\def\affil#1{\affiliation{#1}}

\begin{document}

\title{Spatially focused magnetic hyperthermia: comparison of MRSh and sLLG equations} 

%\author{Zsófia Iszály$^1$, Anna Husztek$^2$, Bujar Mehmeti$^{3,2}$, Zoltán Erdélyi$^2$, Árpád Szöőr$^2$, Miklós Béres$^4$, József Korózs$^4$, Viktória Bacsó$^{1,5}$, István Nándori$^{1,4}$, István Gábor Márián$^{1,2}$}

\author{Zs.~Isz\'aly} 
\affil{HUN-REN Atomki, P.O.Box 51, H-4001 Debrecen, Hungary} 

\author{A.~Husztek}
\affil{University of Debrecen, P.O.Box 105, H-4010 Debrecen, Hungary}

\author{B.~Mehmeti}
\affil{University of Wisconsin-Madison, Department of Medical Physics, 1111 Highland Ave, Madison, WI, 53705, USA}
\affil{University of Debrecen, P.O.Box 105, H-4010 Debrecen, Hungary}

\author{Z.~Erd\'elyi}
\affil{University of Debrecen, P.O.Box 105, H-4010 Debrecen, Hungary}

\author{\'A.~Sz\"o\H{o}r}
\affil{University of Debrecen, P.O.Box 105, H-4010 Debrecen, Hungary}

\author{M.~B\'eres}
\affil{University of Miskolc, Institute of Physics and Electrical Engineering, H-3515, Miskolc, Hungary}

\author{J.~Kor\'ozs}
\affil{University of Miskolc, Institute of Physics and Electrical Engineering, H-3515, Miskolc, Hungary}

\author{V.~Bacs\'o} 
\affil{HUN-REN Atomki, P.O.Box 51, H-4001 Debrecen, Hungary} 
\affil{Ag\'ora, P.O.Box 69, H-4015, Debrecen, Hungary}

\author{I.~N\'andori}
\affil{University of Miskolc, Institute of Physics and Electrical Engineering, H-3515, Miskolc, Hungary}
\affil{HUN-REN Atomki, P.O.Box 51, H-4001 Debrecen, Hungary} 
\affil{University of Debrecen, P.O.Box 105, H-4010 Debrecen, Hungary}

\author{I.~G.~M\'ari\'an}
\affil{HUN-REN Atomki, P.O.Box 51, H-4001 Debrecen, Hungary} 
\affil{University of Debrecen, P.O.Box 105, H-4010 Debrecen, Hungary}

\begin{abstract} 
Magnetic hyperthermia with metallic nanoparticles is a therapeutic strategy that relies on heating cancer cells to levels sufficient to damage or destroy them. After injection, the nanoparticles accumulate in tumor tissues, where they transfer energy from the applied time-dependent magnetic field to the surrounding medium, thereby increasing the local temperature. This heating effect can be spatially focused (superlocalized) by combining AC and DC magnetic fields. Heat generation arises either from the rotation of the particle or from the rotation of its magnetic moment. The theoretical framework is provided by the Martsenyuk-Raikher-Shliomis (MRSh) equation for the former and the stochastic Landau-Lifshitz-Gilbert (sLLG) equation for the latter. However, by using the concept of magnetic and ordinary viscosity, the results of these approaches can be directly compared, which is our goal in this work, with special emphasis on their ability to achieve spatial localization. On the basis of this comparison, we propose the use of perpendicular AC and DC magnetic fields for image-guided thermal therapy with magnetic particle imaging.
\end{abstract}

\pacs{75.75.Jn, 82.70.-y, 87.50.-a, 87.85.Rs}
\maketitle

%------------------------------------------------------------------------------------ 
\section{Introduction}
\label{sec_intro}
%------------------------------------------------------------------------------------
Magnetic nanoparticle hyperthermia is an adjuvant treatment strategy in cancer therapy where injected magnetic nanoparticles (MNPs) are used to increase the temperature locally, see e.g., \cite{PRA1,PRA2,PMB1,PMB2,pankhurst,ortega_d,pankhurst_progress,perigo_e_a,cabrera_d,clinapp_1,clinapp_2,review_mnp}. The capacity of metallic nanoparticles to preferentially accumulate in tumor tissues is the primary benefit of using them in hyperthermia. This is due to the tumor's enhanced permeability and retention (EPR) impact, which enables nanoparticles to penetrate and accumulate in the tumor location, for more details and references, see e.g., \cite{thesis_mehmeti}. The abnormal design of tumor blood vessels, which are leaky and have irregular shapes, 
causes the EPR effect, enabling nanoparticles to travel through the gaps and collect in the tumor tissue. However, there are several difficulties and limitations linked to magnetic nanoparticle hyperthermia. One of the most difficult tasks is optimizing the nanoparticle characteristics and external energy source to obtain the best therapeutic results. This involves finding the best MNPs size, form, surface coating, and concentration, as well as the best energy dosage and frequency.

Let us note, that in terms of the properties of a suitable MNP system, the reduction of the size of the dispersed phase has a beneficial effect up to a certain limit. The larger the size of the dispersed phase, the less negligible the effect of gravity is compared to the interfacial forces, and therefore the separation of the dispersion according to density can be predicted. The density of magnetic solid particles is typically at least 2-3 times greater than that of the liquid phase. In order to reduce the effect of gravity, we need to reduce the size of the solid particles. Based on an order of magnitude calculation, one can see that, on a millimeter scale, the gravitational force and the interfacial forces are of similar magnitude. From a practical point of view, effects below 1$\%$ are usually neglected, so if one goes down another order of magnitude in percentage terms, the order of magnitude of gravity is 0.1$\%$ compared to interfacial forces in the micrometer range which demonstrates that MNP in the nano range, i.e. nanodispersions are preferable. Finally, it is important to note that in magnetic hyperthermia the density of the ferrofluid is assumed to be very small, so, two-particle effects such as discussed in Ref. \cite{astroidcurves_1,astroidcurves_2,PMB3} are usually neglected.

Another challenge is to ensure the safety of the procedure, as overheating can harm healthy tissues and result in adverse effects such as discomfort, inflammation, or tissue necrosis. Thus, there is a need for spatially focused heating, which can be done by using the combination of static gradient (DC) and time-dependent alternating (AC) magnetic fields, where the temperature increase is observed only where the static field vanishes \cite{mpi_test}. The improvement of the efficiency and the spatial focusing ability of MNP hyperthermia receives considerable attention, see e.g.,~\cite{adv_func_mat,review_mnp} and \cite{non_local_heat,focused_hyperthermia,focused_hyperthermia_2,focused_hyperthermia_3,focused_hyperthermia_4,recent_focused_hyperthermia,stat_along_increased}. It is known that the transferred energy is given by a bell-shaped curve as a function of the DC field amplitude, see e.g., Fig.~(10.4) of \cite{review_mnp} or \cite{superlocal_jmmm, superlocal_jpd} and for more details \cite{MRSh_theory_focused}.

The energy transferred in a single cycle of the AC field is identical to the area of the dynamic hysteresis loop, and it is related to the imaginary part of the frequency-dependent susceptibility. Measurements of dynamic hysteresis loops under the influence of AC and DC magnetic fields were reported in several publications, see for example \cite{Mehdaoui,Myrovali,Onodera}. Similarly, a theoretical study of the dynamics of magnetization in the presence of AC and DC fields was carried out, for example in \cite{Dejardin,Mrabti,Murase,MRSh_theory_focused}, and recently, the study of the influence of static magnetic field on the specific absorption rate of the randomly oriented assembly of magnetic nanoparticles was reported in \cite{Rytov}. In addition, the influence of anisotropy and magneto-dipole interaction effects was investigated in the presence of AC and DC fields, see e.g., \cite{usov_claster,Zhao,Usov_2019,viscous_rotating,Engelmann,hip_mc}.

The single-domain MNP dissipates heat as a result of relaxation losses under the applied AC magnetic field. This heat generation occurs due to either the physical rotation of the particle as a whole (Brownian relaxation) or the internal rotation of the magnetic dipole (N\'eel relaxation) while the particle is assumed to be fixed \cite{brown_neel_relax_1,brown_neel_relax_2,brown_neel_relax_3}. Although various mathematical approaches are necessary for a theoretical description of the energy absorption processes of immobilized and freely rotating nanoparticles, the usual theoretical framework is the Martsenyuk-Raikher-Shliomis (MRSh) \cite{mrsh} equation for the Brownian, and the stochastic Landau-Lifshitz-Gilbert (sLLG)  \cite{sLLG} equation for the N\'eel relaxation. 

Spatially-focused heating was studied recently in \cite{superlocal_jpd} employing the sLLG equation and a considerable polarization effect was shown in spatial focusing ability for small frequencies and large field strengths (where linear response theory cannot be applied). In particular, the theoretical results of \cite{superlocal_jpd} predicted a strong spatial focusing ability for the perpendicular combination of DC and AC magnetic fields. However, this result is valid only in the N\'eel regime, thus, one must extrapolate/extend the theoretical considerations to the case of Brownian relaxation in order to be able to use its importance in practice. 

To do that extension, one possible choice is to use the MRSh equation and compare previous results of the sLLG and the MRSh equations directly. Our reasons for this choice are the following. It is well-known that under some constraints, the two different regimes, i.e. the Brownian and the N\'eel relaxations are very much separated by each other because the relaxation time $\tau$ is either dominated by $\tau_{\mr B}$ (Brown relaxation time) or by $\tau_{\mr N}$ (N\'eel relaxation time). So, it is a good approximation to use either the MRSh or the sLLG equations. However, by using the concept of magnetic and ordinary viscosity, the results of the sLLG equation can, in principle, be used to describe not just the motion of the magnetic dipole but also the particle. Of course, in this case the parameters (damping factor, frequency, magnetic field amplitude etc.) must be chosen appropriately, and these are not necessarily the same if one wants to obtain the same results by the two equations. 

Indeed, one of our goals here is to show that one can reproduce well-known dynamic hysteresis loops and bell-shaped spatially-focused energy loss results of the MRSh approach by the sLLG equation. In particular, we demonstrate this by reproducing the MRSh results given on Fig.~2 and Fig.~3 of Ref.~\cite{MRSh_theory_focused} in the framework of the sLLG equation. For example, the spatial focusing ability is different in the right (low frequencies) and left (high frequencies) panels of Fig.~2 of Ref.~\cite{MRSh_theory_focused}. In this work, we show that this is the consequence of the use of a parallel combination of DC and AC magnetic fields and the breakdown of the linear response theory for low frequencies which is signaled by the corresponding dynamic hysteresis loops on Fig.~3 of Ref.~\cite{MRSh_theory_focused}. We can conclude then that for the same reason (lack of the linear response), the perpendicular orientation of DC and AC magnetic fields are in favor in the low-frequency limit both for the MRSh and sLLG cases. 

Let us note that the low-frequency limit receives important application in the so-called image-guided thermal therapy by magnetic particle imaging (MPI), see e.g., \cite{recent_focused_hyperthermia,MRSh_theory_focused,adv_func_mat,review_mnp}. MPI is a tomographic imaging method \cite{gleich} that is based on the magnetization response of MNPs to generate an image. This provides high spatial and temporal resolution with excellent contrast (see e.g., \cite{herz,vaalma}) and its applicability to cancer imaging has also been demonstrated \cite{yu}. The other advantage of the method is its combination with magnetic hyperthermia to provide real-time feedback and to overcome the nonspecific heating problem \cite{MRSh_theory_focused,recent_focused_hyperthermia,mpi_test}. Basically, the same AC magnetic field can be used both for MPI and magnetic heating but with the requirement of low frequencies for MPI.

%--------------------------------------------------------------------------------------
\section{The MRSh equation}
\label{sec_MRSh}
%--------------------------------------------------------------------------------------
In cancer therapy by magnetic nanoparticle hyperthermia, MNPs are injected into the bloodstream during the process and allowed to collect at the tumor site because of the EPR effect. Once the MNPs are in position, they are exerted by an external AC magnetic field causing them to quickly heat up and destroy the cancer cells that encircle them. The application has limitations. The frequency cannot be chosen too high to minimize the eddy currents; a typical maximum value is around $500-1000$ kHz. In addition to that, the product of the frequency and the amplitude of the applied field, which is proportional to the power injected, has an upper bound, the Hergt-Dutz limit \cite{hergt_dutz,hergt_dutz_2,hergt_dutz_3}. Thus, amplitudes are typically in the range of $10-40$ mT $\sim$ $10-30$ kA/m, while frequencies are in the range $100-500$ kHz. Although the diameter of the MNP matters \cite{usov_claster}, we do not consider this effect here and leave it for future work. In general, a good choice is between 10 and 50 nm which we use in our analysis. In addition to that, here we restrict the theoretical study to the isotropic case where no shape and no crystal anisotropy are considered for simplicity. Taking into account anisotropy would be a logical next step, but it is out of the scope of this paper. 

As we argued, single domain MNPs dissipate heat due to Brown (rotation of the particle) or N\'eel (rotation of the magnetic moment) relaxation losses. To decide which process dominates, one has to consider the parameters: e.g., for high frequency ($f \gtrsim 100$ kHz) and small diameter ($\sim 20$ nm), only the orientation of their magnetic moment has to be taken into account \cite{usov_hysteresis,lyutyy_general,neel_brown,viscous_rotating}. 

Calculation of the energy loss in a single cycle requires time-dependent magnetization ${\bf m}$, which can be obtained by solving the phenomenological magnetization equation derived by Martsenyuk, Raikher, and Shliomis, i.e. the MRSh equation \cite{mrsh}. This equation is valid at moderate high field amplitudes and frequencies and is typically used to describe Brownian relaxation,
\begin{gather}
\frac{\rm{d}{\bf m}}{{\rm{d}}t}  = {\bf \Omega} \times {\bf m} 
- \frac{{\bf H} [{\bf H} \cdot ({\bf m} - {\bf m_0})]} {\tau_\parallel H^2}
+ \frac{{\bf H} \times ({\bf m} \times {\bf H})} {\tau_\perp H^2}, \nn
\label{mrsh}
\frac{{\bf m_0}}{m_S}=
L(\xi) \frac{\bf H}{H}, \quad \quad
\xi=\frac{m_S V \mu_0 H}{k_B T},
\end{gather}
where ${\bf \Omega}$ is the flow vorticity of the suspending fluid, ${\bf H}$ is the applied magnetic field, $m_S$ is the saturation magnetization, $L(\xi) \equiv\coth \xi- \frac{1}{\xi}$ is the Langevin function, $V$ is the volume of the magnetic nanoparticle (without coating), $\mu_0$ is the vacuum permeability, $k_B$ is the Boltzmann constant, $T$ is the absolute temperature,
$\tau_\parallel = \frac{ d \ln(L(\xi))}{d \ln \xi} \tau_B$ 
and 
$\tau_\perp = \frac{2 L(\xi)}{\xi-L(\xi)} \tau_B$ 
are the parallel and perpendicular relaxation times related to the Brownian relaxation time $\tau_B$ which depends on the viscosity $\eta_B$ of the suspending fluid, the magnetic diameter $d$, and the thickness $\beta$ of the coating and reads as,
\beq
\label{taub}
\tau_B = \frac{\pi\eta_B(d + 2\beta)^3}{2 k_B T}.
\eeq
It is important to note that the solution of the MRSh equation \eq{mrsh} provides us the time dependence of the average of the individual magnetization vectors.

Assuming an unidirectional alternating magnetic field and no bulk flow, Eq.~\eq{mrsh} can be reduced to a one-dimensional form,
\begin{gather}
    \frac{{\rm{d}} M}{{\rm{d}}t} = -\frac{1}{\tau_B} 
    \left( 1 - \frac{H}{H_e} \right) M, \nn
\label{1dmrsh}
    M=L(\xi_e), \quad \quad 
    \xi_e = \frac{m_S V \mu_0 H_e}{k_B T},
\end{gather}
where the dimensionless magnetization $M=m/m_S$ is used, and $H_e(t)$ called the effective magnetic field is calculated based on the assumption that the magnetization, $M(t)$ at time $t$, would be in equilibrium if the only external field present was $H_e(t)$.
%, thus $H_e(t)$ is defined by the second row of \eq{1dmrsh}.

In this work Eq.~\eq{1dmrsh} is solved to obtain the average of the magnetization vector component $M(t)$ that is parallel to the unidirectional alternating magnetic field. 

The parameters are chosen to be: $\eta_B=10^{-3}$ Pa\,s, ${d=20}$ nm, $\beta=2$ nm, $T=300$ K, $m_S=4.5 \times 10^{5}$ A/m, for an easy comparison with previous works such as \cite{MRSh_theory_focused}, where the same values are given (typical choices for a single crystal Fe$_3$O$_4$ \cite{Fannin}). This yields the Brownian relaxation time $\tau_B\approx 5.25\times 10^{-6} $ s.

%--------------------------------------------------------------------------------------
\section{The stochastic LLG equation}
\label{sec_LLG}
%--------------------------------------------------------------------------------------
Another option to obtain the time-dependent magnetization is the use of the sLLG, which describes the motion of the magnetization vector of a single (isotropic) MNP that undergoes the N\'eel relaxation. The stochastic dynamics has been studied for a long time, supported by experimental works, e.g., \cite{thermal_exp} and summarized in detail in reviews, e.g., Ref.~\cite{thermal_summary}. The sLLG equation \cite{sLLG,sLLG_2,sLLG_3,sLLG_4},
\begin{equation}
\label{sLLG}
\frac{\rm{d}{\bf M}}{{\rm{d}}t}  = -\gamma' [{\bf M \times (H_{\rm{eff}}+H)}] 
+ \alpha' [[{\bf M\times (H_{\rm{eff}}+H)]\times M}],
\end{equation}
where thermal fluctuations are taken into account by the random magnetic field ${\bf H} = (H_x, H_y, H_z)$, where its components are independent Gaussian white noise variables,
\begin{equation}
\label{stochastic_field}
\langle H_i(t) \rangle = 0, \hskip 0.5 cm \langle H_i(t_1) H_j(t_2) \rangle = 2 \, D \, \delta_{ij} \, \delta(t_1 -t_2)
\end{equation}
with $i,j = x, y, z$ and $\gamma' = \mu_0 \gamma_0 /(1+\alpha^2)$, $\alpha' = \gamma' \alpha$ with dimensionless damping $\alpha = \mu_0\gamma_0\eta_N$ with damping factor $\eta_N$ and $\gamma_0 = 1.76 \times 10^{11}$~Am$^2$/Js  is the gyromagnetic ratio, while $\mu_0 = 4 \pi \times 10^{-7}$~Tm/A (or N/A$^2$) is the vacuum permeability. Here ${\bf M} = {\bf m}/m_S$ is a 3-dimensional unit vector, which is normalized by its saturation magnetization, $m_S$. The parameter $D$ corresponds to the fluctuation-dissipation theorem, e.g., in Ref.~\cite{path_int_sllg}, and is defined as $D = \eta_N k_B T/(m_S V \mu_0)$ with the Boltzmann factor $k_B$, the absolute temperature $T$, and the volume of the particle $V$. The angular brackets stand for averaging over all possible realization of the stochastic field, ${\bf H}(t)$, and $\delta(t)$ is the Dirac $\delta$ function. 
This results in the Néel relaxation time,
\begin{equation}
    \tau_N=\frac{(1+\alpha^2)}{2 \alpha \gamma_0} \frac{m_S V}{ k_B T}.
\end{equation}

In order to solve the sLLG equation numerically as an Ito process, it is useful to rewrite it in terms of spherical coordinates. This also ensures that the magnitude of the magnetization remains unchanged. In a simple case of a unidirectional alternating magnetic field the sLLG equation takes the form:  
\begin{align}
\frac{\rm{d}}{\rm{d}t} \phi &= 
-H_{\rm{eff}} \; \gamma' +
\frac{1}{\sin\theta} \sqrt{\frac{1}{2\tau_N}} n_\phi,
\nonumber \\ 
\frac{\rm{d}}{\rm{d}t} \theta &= 
-H_{\rm{eff}} \; \alpha' \sin \theta+
\frac{1}{2\tau_N} \frac{\cos\theta}{\sin\theta} + 
\sqrt{\frac{1}{2\tau_N}} n_\theta,
\end{align}
where the role of random white noise is played by the Gaussian random variables $n_\phi$ and $n_\theta$, which satisfy the relations,
\begin{equation}
\label{nproperties}
\langle n_i(t) \rangle = 0, \quad
\langle n_i(t_1) n_j(t_2) \rangle = 2 \delta_{ij} \, \delta(t_1 -t_2) ,
\end{equation}
with $i,j=\phi, \theta$.
In case of a non-unidirectional effective field, the formulas are similar, but lengthier, thus not presented here. These equations are solved and averaged over a large number of runs to obtain the magnetization vector averaged over all possible realizations of the stochastic field. This is an important difference compared to the MRSh equation, where ${\bf m}$ already stands for the averaged magnetization.

Let us also introduce the frequency-like parameters, $\omega_L =  H \gamma'$ and $\alpha_N = H \alpha'$, where the amplitude of the external magnetic field $H$ is also incorporated. The damping parameter can be chosen as $\alpha = 0.1$ or $\alpha = 0.3$, which are used in Ref.~\cite{Lyutyy_energy} and in Ref.~\cite{Giordano} respectively. In order to enable a direct comparison with the results of MRSh, the value of $\alpha$ is changed in this work. If $\alpha=1.0$, the two models become qualitatively identical. The $m_S$ and $T$ are chosen to be the same as the ones listed in the previous section. However, it is important to note that, due to the different mechanisms, the relaxation times differ ${\tau_N=1.5 \times 10^{-7} \text{s} \neq \tau_B}$. To be in the N\'eel relaxation range, the volume of the magnetic particle must also be different, $V \approx 2.43 \times 10^{-22}$ m$^3$. With these parameters $\omega_L$ and $\alpha_N$ can be obtained.

%--------------------------------------------------------------------------------------
\section{Comparison of the stochastic LLG and the MRSh approaches}
%--------------------------------------------------------------------------------------

%--------------------------------------------------------------------------------------
\subsection{Energy loss (SAR, ILP)}
%--------------------------------------------------------------------------------------

If the solution ($\bf M_{\mr{sol}}$) of the sLLG or the MRSh equation is given, then the volumetric energy loss in a single
cycle can be determined in the following way,
\beq
\label{def_loss}
E = \mu_0 m_S \int_{0}^{\frac{2\pi}{\omega}} {\rm{d}}t 
\left({\bf H}_{\rm{eff}} \cdot \frac{{\rm{d}}{\bf M_{\mr{sol}}}}{{\rm{d}}t} \right).
\eeq
Let us note, that Eq.~\eq{def_loss} contains the average magnetization {\bf $\bf M$}, of a single particle. 
The AC and DC magnetic fields are incorporated in the effective magnetic field $H_{\rm{eff}}$. It is often useful to introduce a dimensionless ratio,
\beq
\label{EpH}
\widetilde E = \frac{E}{2\pi\mu_0 m_S H} \, ,
\eeq
with $H$ being the amplitude of the AC magnetic field. Dividing by the field amplitude and the saturation magnetization, this quantity depends only on the shape of the dynamic hysteresis loop. 

Regardless of the mechanism, if $f$ is the applied frequency, and $E$ is the volumetric energy loss in a single cycle identical to the area of the dynamic hysteresis loop, and related to the imaginary part of the susceptibility, then the product $E \cdot f$ is the volumetric heating power related to the specific loss power (SLP) or specific absorption rate (SAR), which has a dimension of W/g,
\beq
\label{sar}
\frac{E\cdot f}{\phi \, \rho} = \mr{SAR} = \mr{SLP} = \frac{\Delta T \,\, c}{t} \, ,
\eeq
where $\phi$ is the particle volume fraction, $\rho$ is the density of the magnetic particles, $\Delta T$ is the temperature increment, $c$ is the specific heat, and $t$ is the time of the heating period. In addition to that, the intrinsic loss power (ILP) can be derived,
\beq
\label{ilp}
\mr{ILP} = \frac{\mr{SAR}}{H^2 f} \,  {\propto \frac{E}{H^2}} \, .
\eeq
ILP is used to compare SAR values obtained for different values of $f$ and $H$.
In this work $\phi=1$ is used for simplicity, while the density $\rho=5 \times 10^{6}$ g/m$^3$ is chosen, which is a typical value for single crystal Fe$_3$O$_4$ magnetite \cite{Ilg}.

%--------------------------------------------------------------------------------------
\subsection{The MRSh approach}
\begin{figure}[ht] 
\begin{center} 
\includegraphics[width=5cm]{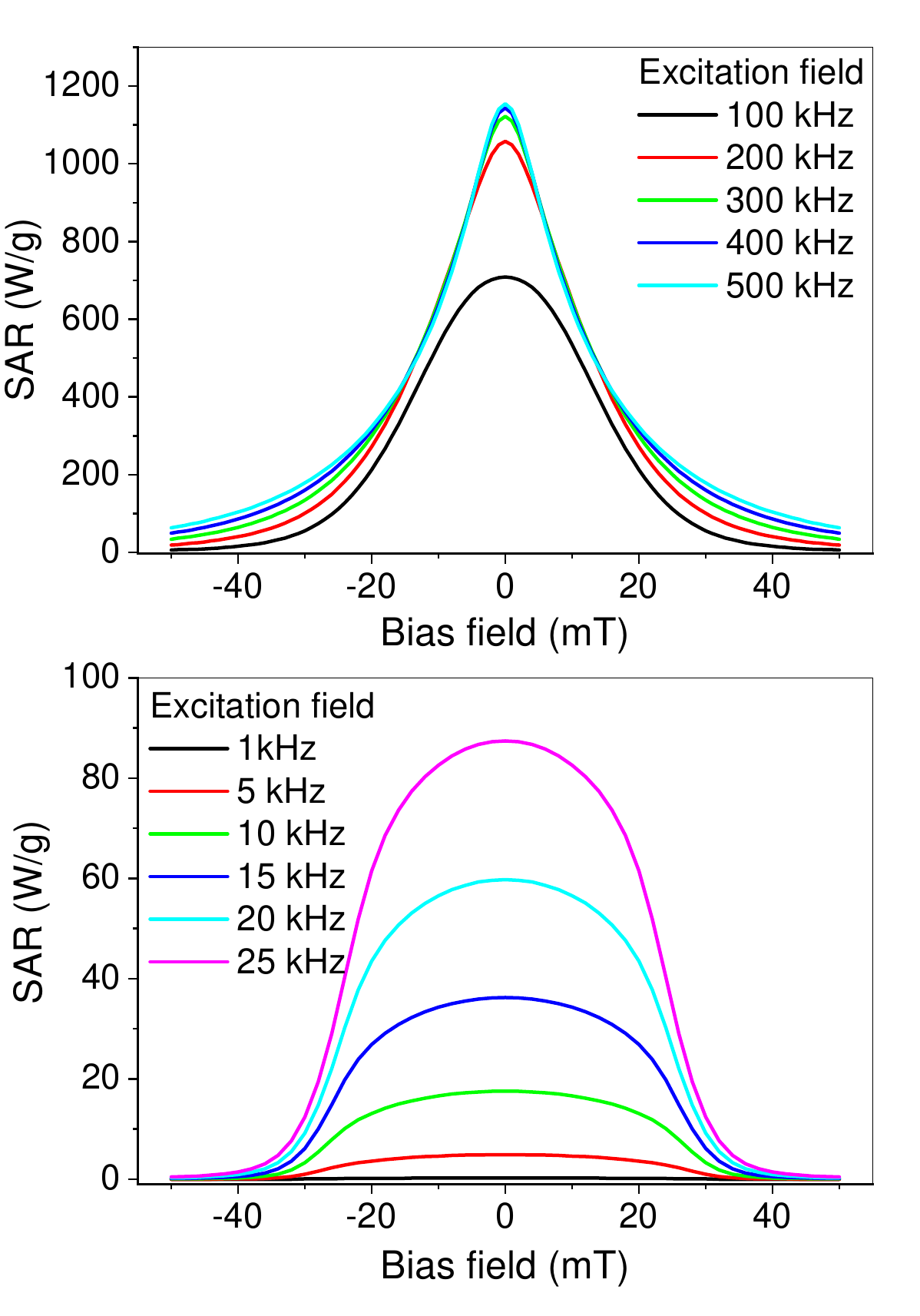}
\caption{(Color online.) 
Spatial focusing is shown by plotting SAR values as a function of the DC field which is in parallel combination of the AC magnetic field. The lower panel is for lower frequencies suitable for MPI and on the upper panel one finds higher frequencies needed for magnetic hyperthermia. The AC field strength is chosen to be 30 mT $\sim$ 24 kA/m.
\label{fig1}
} 
\end{center}
\end{figure}

%--------------------------------------------------------------------------------------
In this subsection, we summarize our results where the MRSh approach is applied to study the spatial focusing ability given for the 
parallel combination of AC and DC magnetic fields.
These results are in complete agreement with previous studies on spatial focusing
\cite{recent_focused_hyperthermia,MRSh_theory_focused,adv_func_mat,review_mnp}. 
The applied magnetic field,
\bea
\label{H_osc_parallel_def}
{\bf H}_{\rm{eff}} = H \, \Big(\cos(\omega t) + b_0,\,\,  0, \,\, 0\Big), \nonumber
\eea
where $H$ and $H b_0$ stands for the amplitude of the applied and the static fields, the latter is also called the bias field.
Based on the numerical solution of the MRSh equation, the SAR can be calculated and in \fig{fig1} (which is directly comparable to Fig.~2 in \cite{MRSh_theory_focused}), one finds the variation as a function of the magnitude of the static field $b_0$. 
%
% Figure 1
%
Note that frequencies and amplitudes typically used in MPI are in the order of 1-25 kHz and around 30 mT $\sim 24$ kA/m, but magnetic hyperthermia requires much larger frequencies (100-500 kHz). As a consequence, the SAR values obtained in MPI are much lower than those found at frequencies used in magnetic hyperthermia. Therefore, MNP heating during an MPI scan is expected to be negligible.

\begin{figure}[ht] 
\begin{center} 
\includegraphics[width=8.6cm]{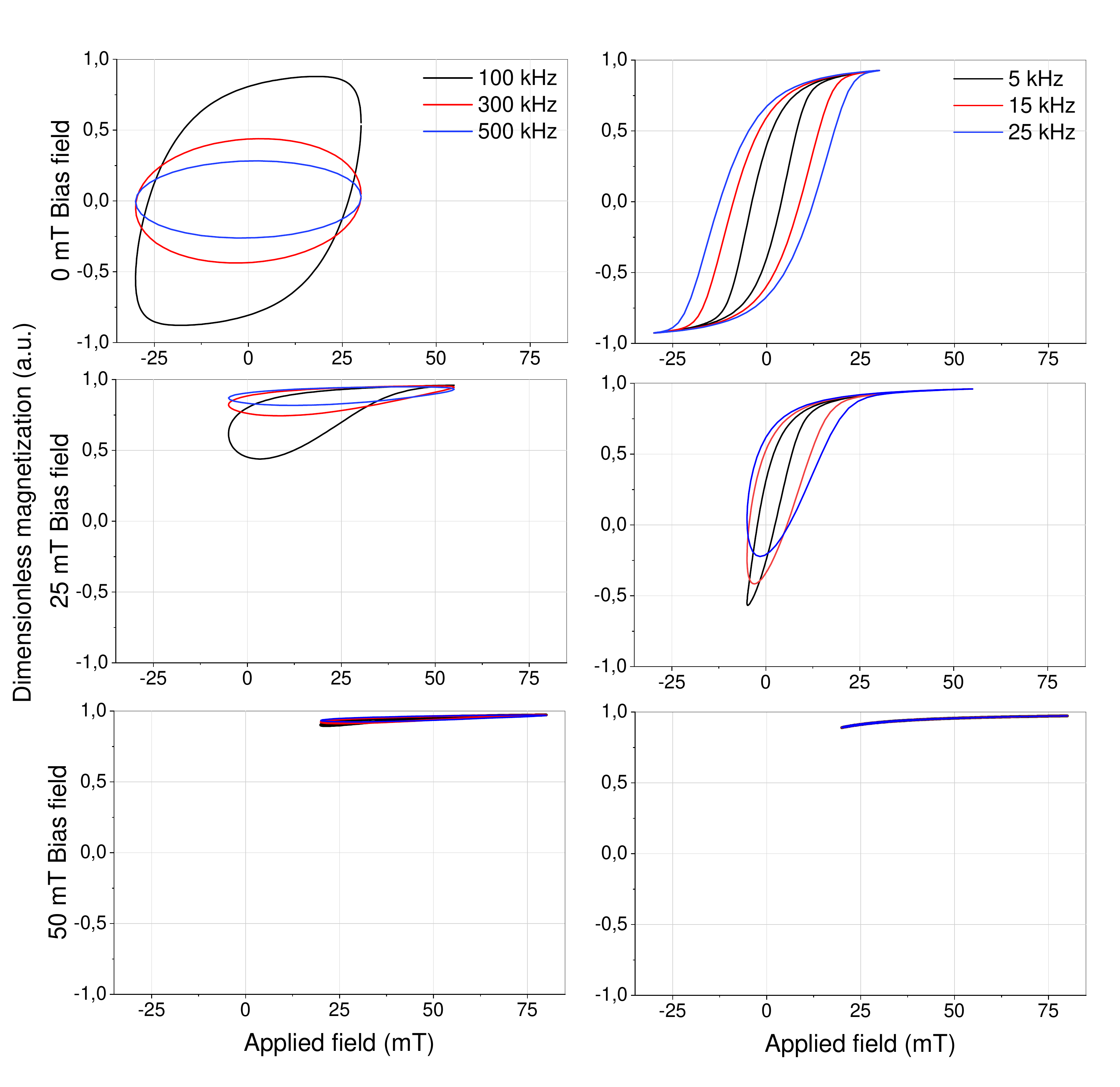}
\caption{(Color online.) Dynamical hysteresis loops are given for various angular frequencies of the applied AC magnetic field while its amplitude (field strength) is fixed. The left column stands for larger frequencies needed for magnetic hyperthermia and on the right column one finds smaller frequency values used in MPI.
The static DC field amplitude is increased from top to bottom.
\label{fig2}
} 
\end{center}
\end{figure}

The spatial focusing ability can be understood by the dynamic hysteresis loops which are given in \fig{fig2} (which is also in complete agreement with Fig.~3. in \cite{MRSh_theory_focused}).
%
% Figure 2
%
The area of the loop decreases with increasing DC field amplitude and drops to zero when the DC field strength exceeds the AC field (see lower panels of \fig{fig2}). This explains the drop in the SAR values in \fig{fig1} with increasing DC field strength. The largest area is found for vanishing DC amplitude, yielding a peak in the SAR values. However, the decrease is much faster for large frequencies compared to low frequencies, i.e., one can observe a broadening of the peak for MPI frequencies. This is because at large frequencies the dynamic hysteresis loops have a much-reduced area, compared to that at the MPI case for the same bias field.

In addition, the DC field also modifies the shape of the hysteresis loop. Before the magnetization reaches the saturation, it takes a teardrop shape, and of course shifts the entire loop horizontally. This is because the resultant applied magnetic field is shifted towards positive values by the magnitude of the DC field, thus causing the magnetization to be asymmetric, or in case of a strong bias field (stronger than the amplitude of the AC field) to remain in saturation. Moreover, increasing the frequency leads to the opening of the hysteresis loop. At very high frequencies, the MNPs cannot follow the applied AC field, causing a delay in their magnetization response and thus in the opening of the dynamic hysteresis loop.

%--------------------------------------------------------------------------------------
\subsection{The stochastic LLG approach}
%--------------------------------------------------------------------------------------
In this subsection, our main goal is to recover the results of \fig{fig1} and \fig{fig2} by using not the MRSh but the sLLG equation for parallel combination of AC and DC magnetic fields,
\bea
{\bf H}_{\rm{eff}} = H \, \Big(\cos(\omega t) + b_0,\,\,  0, \,\, 0\Big). \nonumber
\eea
This is possible because the aforementioned concept of magnetic and ordinary viscosity. However, the value of the damping parameter is clearly different in these two limiting cases, so one has to use different amplitudes and frequencies for the AC field in the sLLG approach with a typical damping suitable for the N\'eel relaxation compared to the MRSh results of \fig{fig1} and \fig{fig2} where the damping is chosen to be well fitted to Brownian relaxation. 
The difference in frequency between the two approaches accounts for the difference in relaxation times, while the difference in the amplitude of the effective field is basically compensating the different volumes of the magnetic nanoparticles.

\begin{figure}[ht] 
\begin{center} 
\includegraphics[width=5cm]{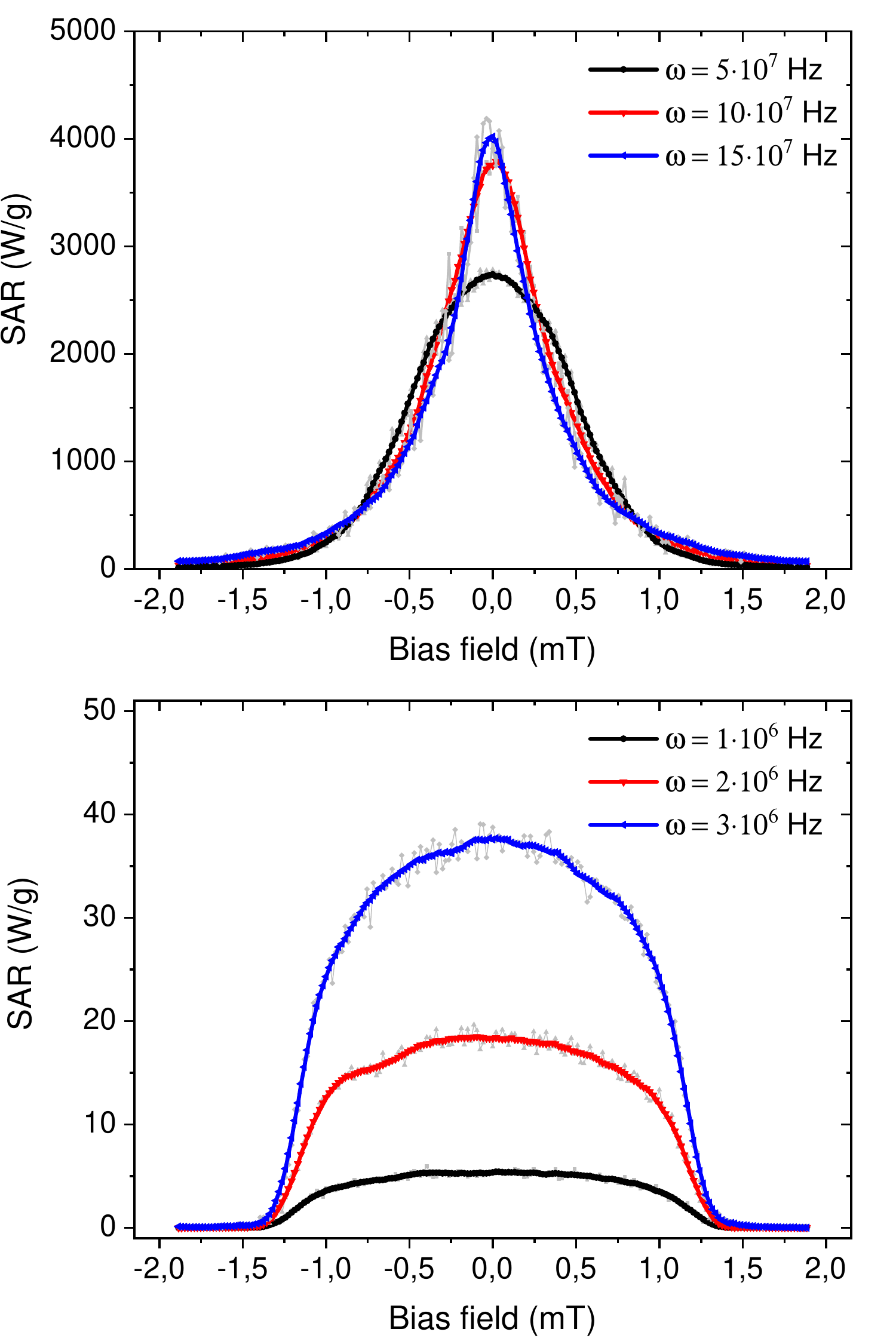}
\caption{Results obtained by the sLLG equation (for field strength 1 kA/m $\sim$ 1.3 mT  and frequencies  $5 \times 10^7$ Hz  to the top and $10^6$ Hz to the bottom). This can be compared to \fig{fig1} given by the MRSh approach. One finds similar bell-shapes for the SAR/ILP values. 
\label{fig3}
} 
\end{center}
\end{figure}

In \fig{fig3} we present the results of our sLLG approach, which can be directly compared to \fig{fig1}. 
%
% Figure 3
%
Similar bell shapes are found for the SAR/ILP values given by the two different methods.
We would like to draw the reader's attention to the broadening of the peak for smaller frequencies used in MPI, see the lower panel of \fig{fig3}.  

\begin{figure}[ht] 
\begin{center} 
\includegraphics[width=9.2cm]{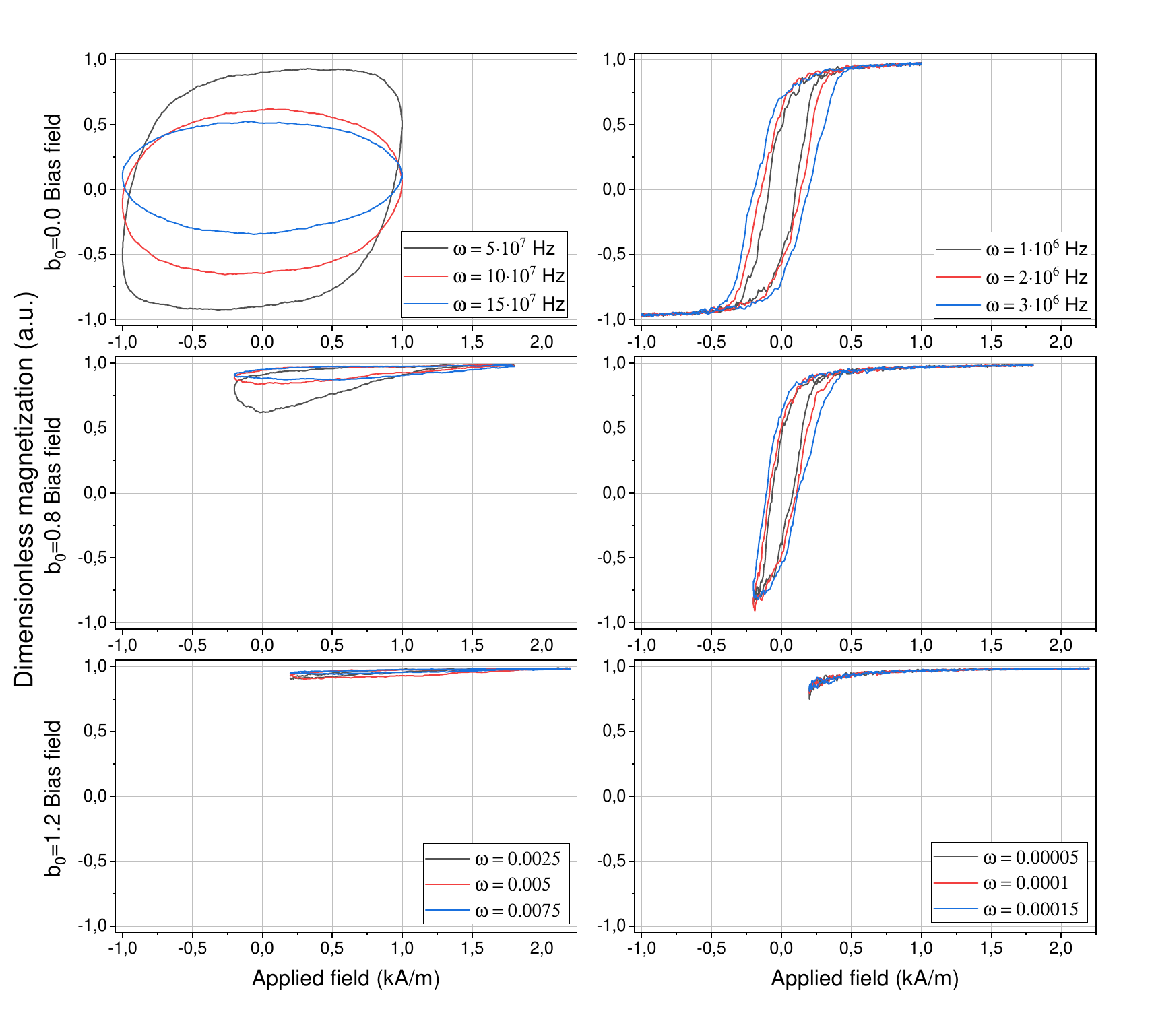}
\caption{(Color online.) Dynamical hysteresis loops obtained by the sLLG equation which are similar to the dynamical hysteresis loops of \fig{fig2} given by the MRSh method. 
\label{fig4}
} 
\end{center}
\end{figure}

The dynamic hysteresis loops obtained by the sLLG equation for various angular frequencies but at a fixed field strength are presented in \fig{fig4}. This figure is the analogue of \fig{fig2} where the MRSh approach was used.
%
% Figure 4
%

The main message is that one can recover the MRSh results using the sLLG method, which is clearly demonstrated by \fig{fig3} and \fig{fig4}.

%--------------------------------------------------------------------------------------
\section{Spatial localization ability}
\label{sec_spatial_loc}
%--------------------------------------------------------------------------------------
In this section we use the conclusion of the previous part, that the sLLG approach can be used (with the appropriate choice of parameters) to recover the results from the MRSh equation. Thus, here we use the sLLG equation to demonstrate that better spatial focusing can be achieved for image-guided thermal therapy by MPI in the perpendicular (instead of the parallel) combination of AC and DC fields,
\bea
\label{para}
\mr{parallel}: \hskip 0.2cm   {\bf H}_{\rm{eff}} = H \, \Big(\cos(\omega t) + b_0,\,\,  0, \,\, 0\Big), \hskip 0.4cm \\
\label{perp}
\mr{perpendicular}: \hskip 0.2cm  {\bf H}_{\rm{eff}} = H \, \Big(\cos(\omega t),\,\,  b_0, \,\, 0\Big). \hskip 0.4cm
\eea

\begin{figure}[ht] 
\begin{center} 
\includegraphics[width=5cm]{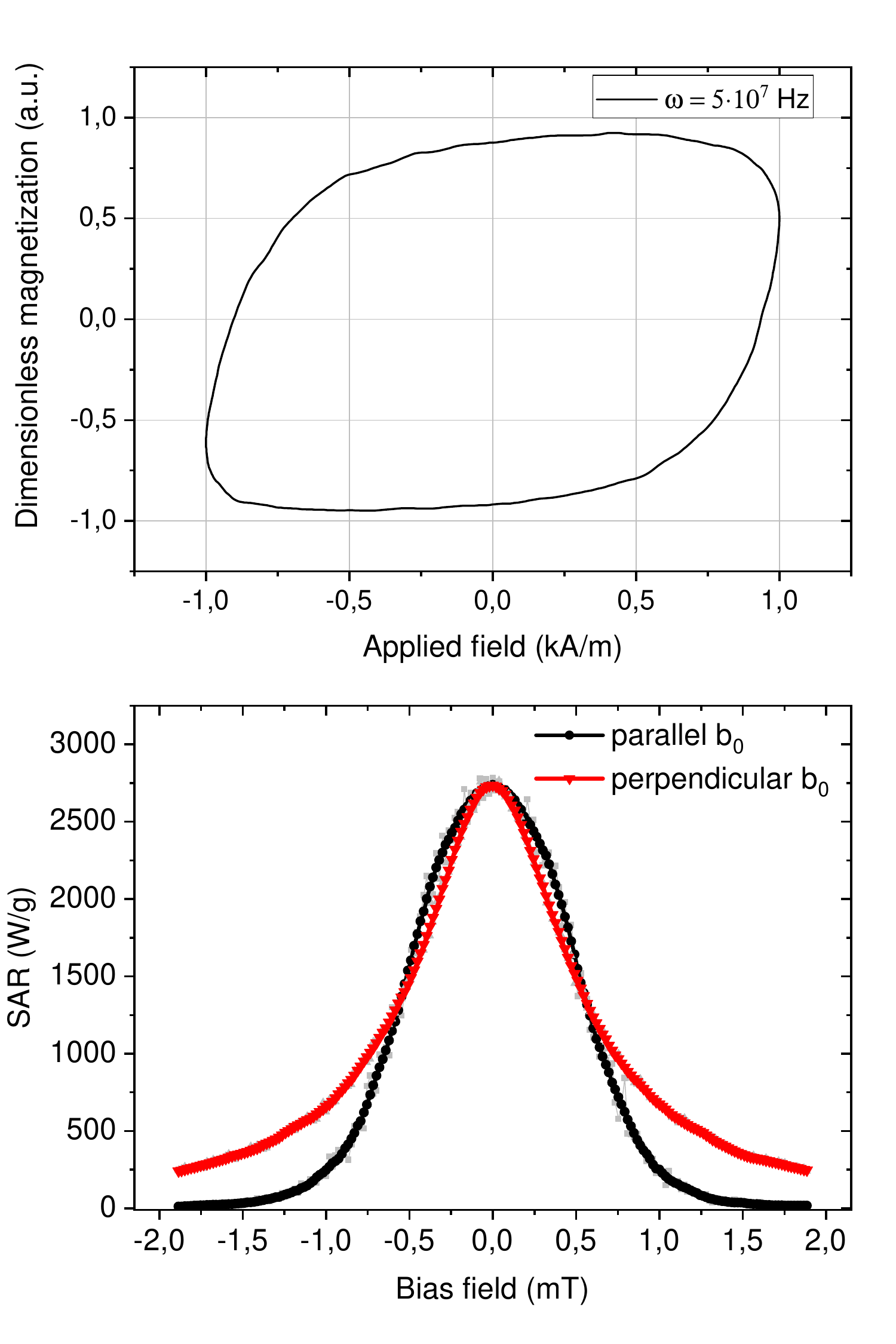}
\caption{(Color online.) Dynamic hysteresis loop (for parallel combination \eq{para}) and spatial focusing of SAR values for parallel \eq{para} and perpendicular \eq{perp} combinations) obtained by the sLLG method for field strength 1 kA/m $\sim$ 1.3 mT and frequency  $5 \times 10^7$ Hz. One finds no difference in the spatial focusing ability of parallel and perpendicular orientation of AC and DC fields.
\label{fig5}
} 
\end{center}
\end{figure}

In \fig{fig5} we show the dynamic hysteresis loop and spatial focusing obtained for large frequencies.
%
% Figure 5
%
For large frequencies, one finds no difference in the spatial focusing ability of parallel and perpendicular orientations of AC and DC fields.

\begin{figure}[ht] 
\begin{center} 
\includegraphics[width=5cm]{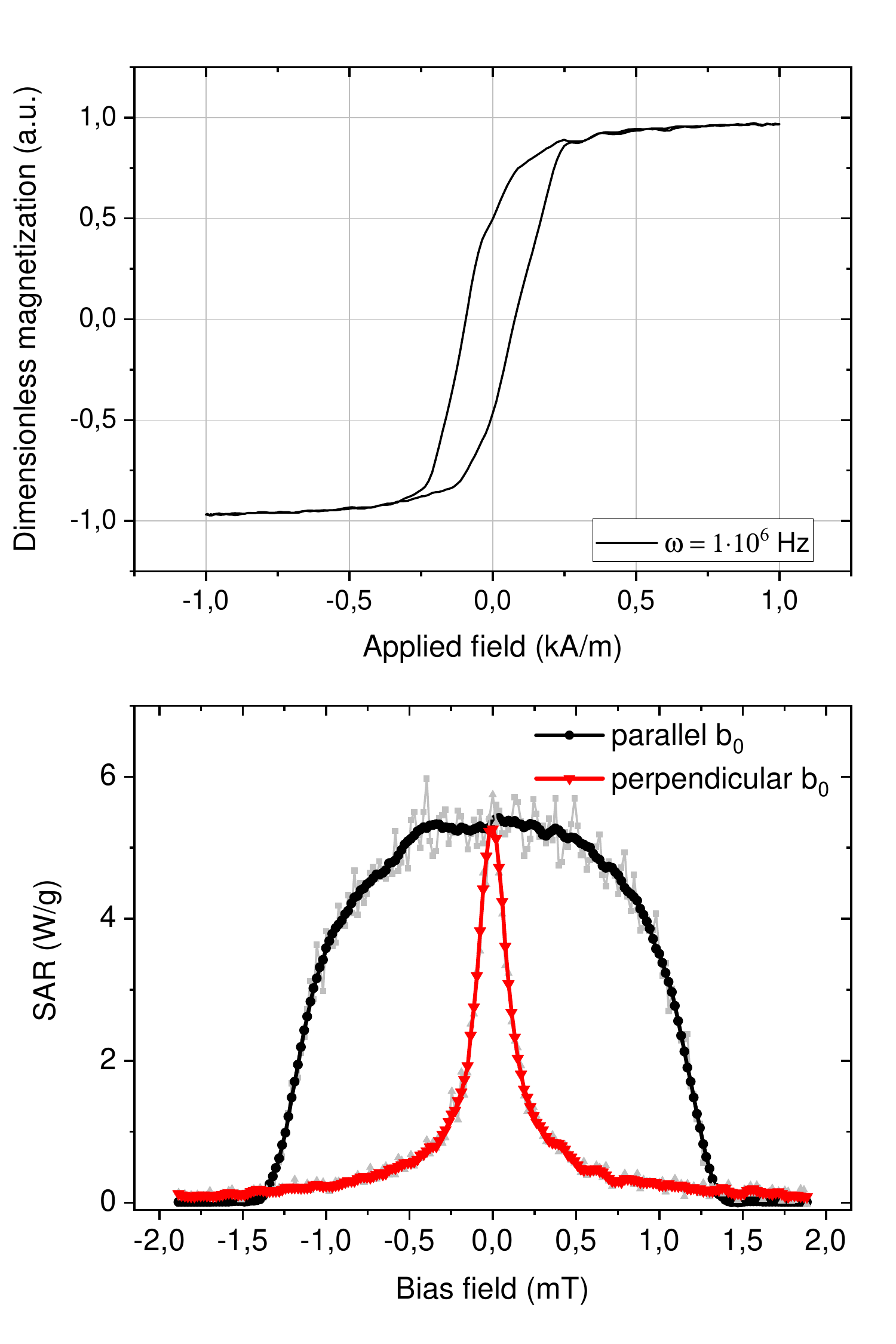}
\caption{(Color online.) The same as \fig{fig5} but for the frequency $10^6$ Hz. It is clearly shown by the figure that one finds a better spatial focusing for perpendicular orientation of AC and DC fields.
\label{fig6}
} 
\end{center}
\end{figure}

In \fig{fig6} we plot the dynamic hysteresis loop and spatial focusing obtained
for small frequencies.
%
% Figure 6
%
For small frequencies, the perpendicular orientation of AC and DC fields results in better spatial focusing ability. This has importance for MPI because of the analogy discussed in the previous section, one expects the same behavior for the Brownian relaxation regime, described by the MRSh equation.

The increased spatial focusing ability of perpendicular (compared to the parallel) orientation of AC and DC magnetic fields can be explained as follows. It was shown in Ref. \cite{superlocal_jpd} that the linear response theory is an adequate approximation for relatively large frequencies and small magnetic fields. Accordingly, linear response theory is not applicable at small frequencies (and at large field strengths). 

Similar conclusions can be drawn from dynamic hysteresis loops. The shape and area of the loops depend on the parameters of the applied magnetic field. The higher the frequency, the more the linear response is valid and the shape of the loop becomes ellipse-like, see the upper panel of \fig{fig5}. For low frequencies, one expects the breakdown of the linear response, and the dynamic hysteresis loop deviates from the ellipse-like, see the upper panel of \fig{fig6}.

For low frequencies, the magnetization vector follows the external magnetic field with almost no phase shift, so energy loss can only be observed if the vector sum of AC and DC fields vanishes. For parallel combination, the magnitude of the DC field should be smaller than or identical to the AC field to find a zero vector sum. However, for perpendicular orientation, even a very small DC field is sufficient to result in a non-vanishing vector sum of AC and DC magnetic fields. 

\begin{figure}[ht] 
\begin{center} 
\includegraphics[width=7.2cm]{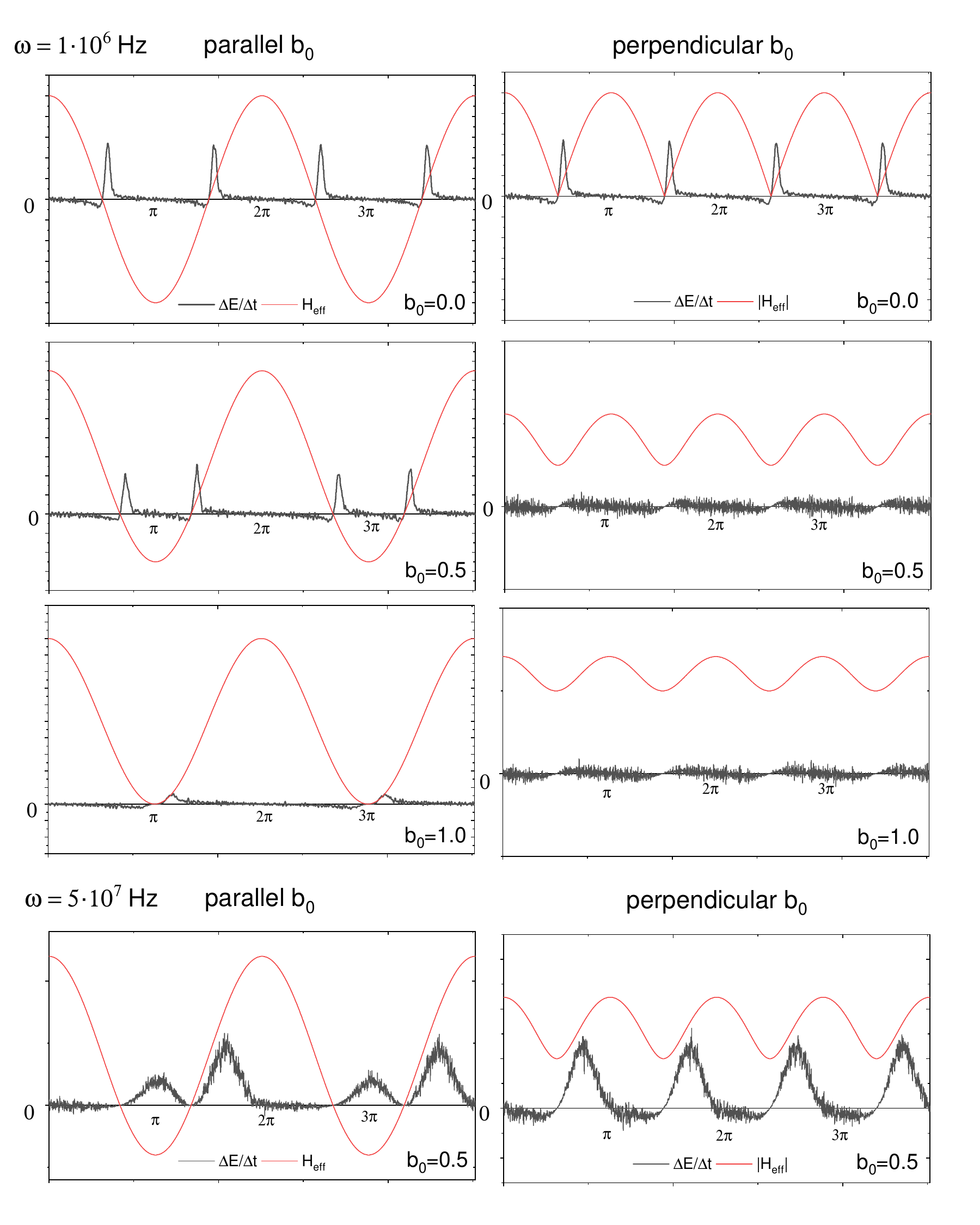}
\caption{(Color online.) Change in energy loss and the applied field versus time, i.e., at different phases over two cycles in parallel and perpendicular combination of the AC and DC field for small frequencies (with three different $b_0$ value) to the top and large frequency to the bottom. 
\label{fig7}
} 
\end{center}
\end{figure}

This energy dissipation mechanism for the vanishing vector can be made more visible and understandable by \fig{fig7}. This shows the change in energy loss over two cycles for both (parallel and perpendicular) cases.  The applied field is also indicated (in the perpendicular combination, the absolute value is shown) to illustrate that the majority of the energy transfer occurs when the sum of the AC and DC fields is zero. For low frequencies, it can happen just for $b_0=0.0$ in the perpendicular case.

%
% Figure 7
%

Therefore, in the perpendicular case, the spatial focusing ability is much stronger than in the parallel case, see the lower panel of \fig{fig6}. If the frequency is high and the field strength is small, one expects a linear response and almost identical spatial focusing ability for the two orientations, see the lower panel of \fig{fig5} and the lower panel of \fig{fig7}, where at a value of $b_0=0.5$, energy loss occurs in both cases.

%--------------------------------------------------------------------------------------
\section{Summary}
\label{sec_sum}
%--------------------------------------------------------------------------------------
In this work, we compared the MRSh and sLLG equations by employing the concept of magnetic and ordinary viscosity with a particular focus on their ability to achieve spatial localization. We demonstrated that the well-known dynamic hysteresis loops and bell-shaped spatially-focused energy loss results obtained from the MRSh approach can be reproduced by the sLLG equation with appropriate choice of parameters. 

\begin{figure}[ht] 
\begin{center} 
\includegraphics[width=5cm]{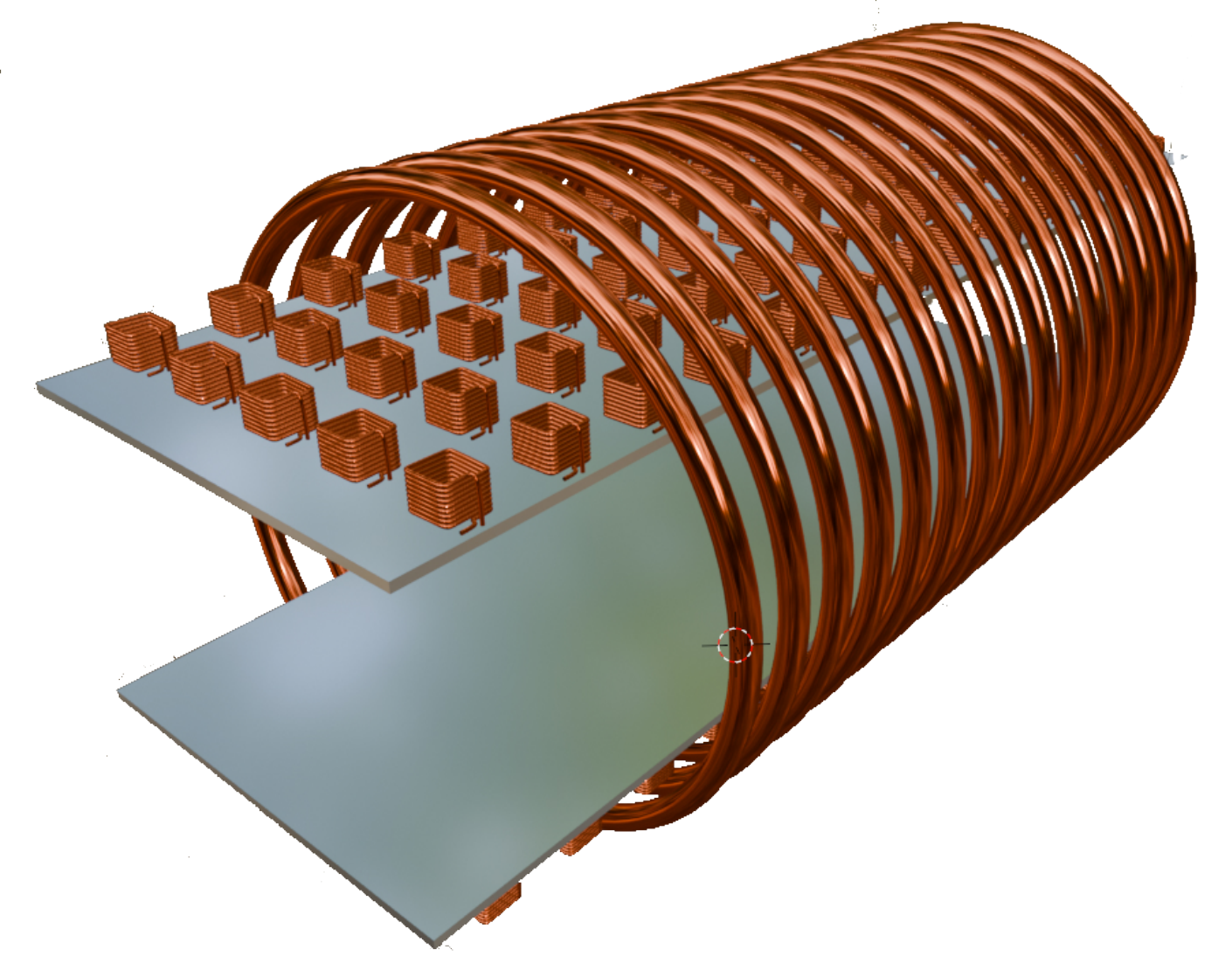}
\caption{(Color online.) Schematic diagram of the concept of equipment directly applicable in medicine.
\label{fig8}
} 
\end{center}
\end{figure}

Thus, it is likely that the results obtained with sLLG are true regardless of the relaxation mechanism.
This confirms our previous finding that, in the case of low-frequency and high-amplitude magnetic fields, perpendicular static and variable magnetic fields provide significantly better spatial focusing than parallel arrangements. Based on our results, we therefore recommend the use of perpendicular combination of alternating and static magnetic fields in MPI-based image-guided magnetic hyperthermia, which requires low frequency and high field strength. Our future goals include supporting spatial localization with the MRSh equation and experimental verification. Further development of the setup used for experimental verification could lead to a device that can be used directly in medicine, the concept of which is shown in the schematic diagram \fig{fig8}.
%
%
% Figure 8
%

\section*{Acknowledgement}
The support for the CNR/MTA Italy-Hungary 2023-2025 Joint Project "Effects of strong correlations in interacting many-body systems and quantum circuits" is gratefully acknowledged.
Supported by the EK\"OP-25-2 University Research Scholarship Program of the Ministry for Culture and Innovation from the source of the National Research, Development and Innovation Fund.
Project no. TKP2021-NKTA-34 has been implemented with the support provided from the National Research, Development and Innovation Fund of Hungary, financed under the TKP2021-NKTA funding scheme. Supported by the University of Debrecen Program for Scientific Publication.

\FloatBarrier

\end{document}